\documentclass[12pt,letterpaper,titlepage]{article}
 \usepackage{color}
 \usepackage{graphics}
\begin{document}
\vspace*{0.2in}
 \begin{center} \Large 
 Single-Particle Excitations in a \\ 
 Two-Dimensional Strong-Coupling Superconductor 
  \end{center}
\vspace*{0.2in}
 \begin{center} 
 {\em D.W. Hess} \\
 Complex Systems Theory Branch \\
 Naval Research Laboratory \\
 Washington, D.C. \ \ 20375-5345
 \end{center}
\vspace*{0.05in} 
 \begin{center}
  {\em J.J. Deisz} \\
  Department of Physical Sciences\\
  Black Hills State University \\
  Spearfish, SD \ \ 57799-9051 \\
 \end{center}
\vspace*{0.05in} 
 \begin{center}
  and {\em J.W. Serene} \\
   Department of Physics \\
   Georgetown University \\
   Washington, D.C. \ \ 20057-0995

\vspace*{0.20in}

{\small {\it To appear: Phil. Mag. Lett.}}

 \end{center}

\vspace*{0.15in}

\begin{center}
 \parbox{4.5in}{{\bf Abstract:}
 We present calculations of the single-particle
 excitation spectrum for a 2D strong-coupling 
 superconductor in a conserving approximation.  
 Spectral weight at low frequency is substantially 
 reduced as the superconducting transition is 
 approached from the normal state. The suppression 
 of low-energy excitations is a consequence of 
 Cooper-pair fluctuations that are described
 self-consistently in the fluctuation exchange
 approximation.  The static and uniform 
 electromagnetic response provides a measure 
 of the super fluid density and a fully 
 self-consistent indication of the superconducting 
 transition temperature. 
 }
\end{center}

  The pseudogap observed in the normal state of the 
  high-temperature superconductors has focused 
  interest on fluctuations directly associated with 
  superconductivity (Emery and Kivelson 1995, Randeria, 
  {\em et al.} 1992, Doniach and Inui 1990). The occurrence 
  of superconductivity in two-dimensional copper-oxide 
  layers encourages speculation that Cooper-pair 
  fluctuations affect the low-energy excitation spectrum 
  irrespective of the pairing mechanism or of the symmetry of 
  the order parameter.  It has been proposed that fluctuations 
  of the phase of the order parameter suppress the physical 
  transition temperature, {\em i.e.} the temperature at which 
  macroscopic signatures of phase coherence first appear, 
  below the temperature at which a gap forms in the single-particle 
  excitation spectrum (Emery and Kivelson 1995, 
  Trivedi and Randeria 1995).  Here we investigate the excitation 
  spectrum in a self-consistent calculation that includes 
  fluctuations of both the amplitude and phase of the order parameter.

  The short coherence lengths characteristic of the
  high-temperature superconductors suggest that an appropriate 
  description lies somewhere between the BCS picture and 
  Bose condensation of bound fermion pairs (Randeria {\em et al}. 1989). 
  The experimental observation of what is apparently a well defined 
  Fermi surface in the high-temperature superconductors and the 
  presence of a sharp drop in $n({\bf k})$ for $|U| \sim W/2$ 
  in quantum Monte Carlo calculations on the Hubbard
  model with an attractive two-body interaction (Trivedi and Randeria 
  1995), indicate that a pairing interaction of intermediate coupling 
  strength on the scale of a typical electronic bandwidth may be of 
  physical interest.\footnote{By contrast, for a sufficiently strong
  pairing interaction, all electrons are expected to participate 
  in the formation of `dielectronic molecules' and no Fermi surface is
  expected.} Recent angle-resolved photoemission experiments on BSCCO 
  have been interpreted as being consistent with a large 
  Fermi surface and therefore a high electronic density 
  (Ding {\em et al.} 1997).  The nature of the crossover 
  from the overlapping Cooper pairs of BCS theory to 
  a Bose condensation of preformed pairs has been explored by 
  several authors in the context of continuum and lattice models 
  (Leggett 1980, Nozi\`{e}res and Schmitt-Rink 1985, Eagles 1969, 
  Randeria {\em et al.} 1992, Luo and Bickers 1993, 
  Micnas {\em et al.} 1990, and S\'{a} de Melo {\em et al.} 1993).
  Taken together,  these works highlight the importance 
  of collective behavior in the particle-particle channel, 
  which is expected to play a significant role, 
  particularly for large interaction strength, in determining 
  the transition temperature, the nature of the phase transition, 
  and, of central interest here, the electronic excitation spectrum.

  In a previous work (Deisz {\em et al.} 1998a), hereafter called DHS,
  we explored superconductivity in the fluctuation exchange approximation (FEA)
  (Bickers {\em et al.} 1989), a conserving approximation (Baym 1962) beyond 
  mean field theory. The superfluid density, its temperature derivative, 
  and the specific heat all show dramatic effects of fluctuations, and scale 
  slowly with increasing lattice size.  This slow scaling so far precludes 
  an unambiguous determination of the type of phase transition, and could
  even be consistent with a Kosterlitz-Thouless transition, but in any
  case it is of interest to know how Cooper-pair fluctuations affect the 
  electronic excitation spectrum in a theory in which fluctuations are powerful 
  enough to correctly renormalize the transition temperature from mean-field
  theory.  To address this question, we calculate single-particle spectral
  functions in the FEA and show the temperature evolution of the 
  excitation spectrum {\em side-by-side} 
  with the superfluid density calculated on lattices of $128 \times 128$ 
  momentum points.  We consider a density $n=0.75$ roughly consistent with 
  that obtained in electronic structure calculations.  This provides
  a connection with the discussion of the nature of the transition presented 
  in DHS which also shows that transition temperatures obtained on small 
  lattices at this density are in good agreement with quantum Monte Carlo 
  calculations. While it is expected that phase fluctuations are 
  most important for the suppression of long-range order, 
  the FEA contains fluctuations of both the amplitude and phase of the 
  order parameter. 

  The central results of this paper are contained in Fig. 1 
  which shows the electronic spectral weight accessible to 
  a photoemission experiment at four temperatures. 
  Spectral weight is indicated by color in an energy ($y$-axis) 
  momentum ($x$-axis) plane.  The momenta shown are along
  the $(1,0)$ direction from $\Gamma$ to $X$ in the square Brillouin 
  zone.  Red indicates the largest spectral weight; spectral weight 
  decreases from red descending in the order of the colors of the 
  spectrum to the smallest indicated by indigo. 
  {\em  Only for the lowest temperature (Panel 1) is there a non-zero superfluid density;
  it is very small compared to that for zero temperature.}  
  The spectrum for the highest temperature, shown in Panel 4, is consistent
  with a band of short-lifetime excitations broadened by correlations and by
  thermal fluctuations. As the temperature is lowered, narrow yellow and red 
  regions appear in the electronic excitation spectrum 
  for some momenta above and below the Fermi surface. 
  The `sharpening' of these single-particle excitations,\footnote{To be
  more precise, by `sharpening' we mean that the peak in the 
  spectral functions for 
  these momenta becomes narrower and spectral weight at the maximum 
  increases.} signals the evolution and emergence of quasiparticle 
  excitations\footnote{While these excitations 
  are relatively `sharp,' the temperature dependence of 
  their lifetime is not consistent with expectations for a Fermi liquid. 
  This may not be surprising given the high temperature and the relatively 
  large energy of these excitations.  In any event, this is not the 
  central point of this paper, and will not be pursued further here.} 
  with energies above and below the chemical potential $\mu$.
  For momenta nearer to the Fermi surface, excitations do not `sharpen' 
  and the light green color indicates that spectral weight at low-energy
  {\em decreases} as the temperature
  is lowered. As we explicitly show below for a point very near the Fermi 
  surface in the $(1,0)$ direction, spectral weight at zero energy  
  is suppressed significantly as the transition is approached; relative to the 
  `sharp excitations'
  the maxima of the spectral functions associated with low-energy excitations 
  are reduced by approximately $50 \%$.  Our results 
  suggest a simple model discussed below.

 \begin{figure}[h]
 \centerline{ \resizebox{4.0in}{4.0in}{\includegraphics{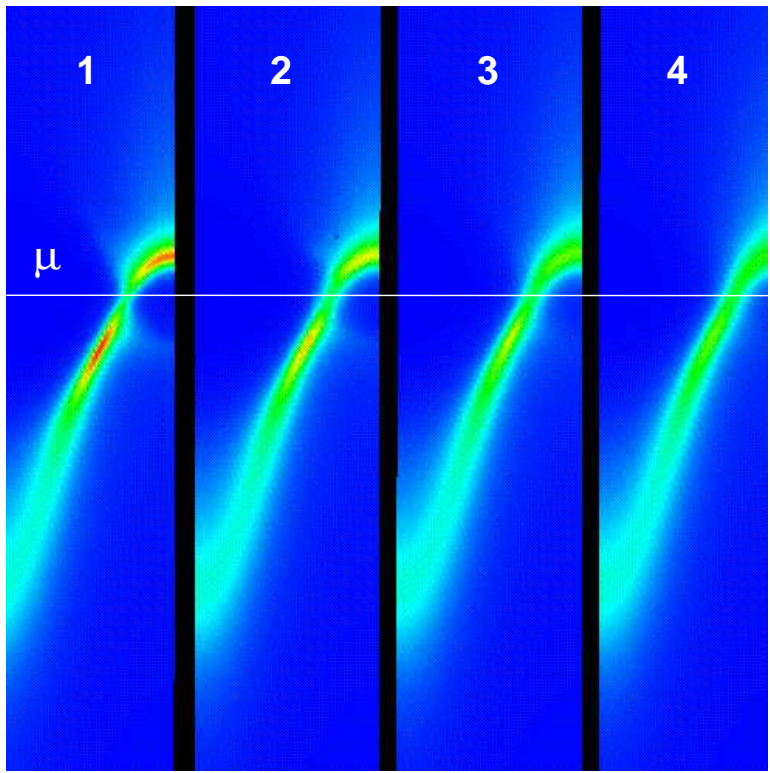}}}
  \caption[]{ {\footnotesize
  Single-particle spectral weight represented by color with
  darkest blue (indigo) corresponding to zero and red to $\sim 1.4$.
  Spectral weight increases in the order of the colors of the rainbow.
  Panel 1 is for $T=0.12$; temperature increases by $0.02$ going
  from panel to panel from left to right. The $x$-axis in
  each panel indicates momentum from $\Gamma$- to $X$-points along
  the $(1,0)$ direction and the $y$-axis indicates energy from
  $-5.5$ to $2.0$. The white line indicates the position of the
  chemical potential. Spectral weight at low energy above the
  transition is suppressed with decreasing temperature (from $4$ to $1$)
  as indicated by the light green region sandwiched between the
  evolving red peaks. Each panel corresponds to an arrow in Fig. 2.
  Here, as in all figures, $U = -4$ and the density is 0.75.
  }}
 \end{figure}

  We begin with a model of coupled Cooper-pair fluctuations and electrons
  that consists of the Hubbard Hamiltonian with an attractive interaction $U <0$
  threaded by a flux $\phi_{B}$, 
\begin{eqnarray}
H(\phi_B)& = & \; -t \; \sum_{{\bf r},\sigma} \;
[ \exp ({2 \pi i {\phi_B }\over{\phi_o L}}) \; c^{\dagger}_{{\bf r},\sigma}
c_{{\bf r} + \hat{{\bf x}},\sigma} + h.c.] \nonumber \\
&& -t \; \sum_{{\bf r},\sigma} \; [c^{\dagger}_{{\bf r},\sigma}
c_{{\bf r} + \hat{{\bf y}},\sigma} + h.c.]
+ U \sum_{{\bf r}} n_{{\bf r},\uparrow} n_{{\bf r},\downarrow},
\end{eqnarray}
 together with the FEA.
 Here $-t$ is the nearest-neighbor hopping 
matrix element, $U (<0)$ is an on-site attractive interaction 
and $\phi_o = hc/e$.

  To self-consistently identify a superconducting phase transition 
  in this model,  we calculated the internal energy $E$, 
  and free energy $F$, as a function of an applied flux
  using the self-consistently determined self-energy 
  and fully renormalized propagator. 
  A finite superfluid density $D_s$ leads to a non-zero curvature in 
  $E(\phi_B)$ and $F(\phi_B)$ at $\phi_{B}=0$
 (Yang 1962, Fisher {\em et al.} 1973, Scalapino {\em et al.} 1992),
\begin{eqnarray} 
 F(\phi_B) & = & F(0) + {1 \over 2} \; D_s(T) \;
  (\frac{\phi_B}{\phi_0})^2 + \cdots , \label{expandF}  \\
 E(\phi_B) & = & E(0) + {1 \over 2} \;
 [D_s - T \frac{dD_s}{dT}] \; (\frac{\phi_B}{\phi_0})^2 + \cdots \; .
\end{eqnarray}
  This procedure is equivalent to obtaining $D_s(T)$ from the 
 long-wavelength and zero-frequency limit of the {\em fully self-consistent}
 electromagnetic response function. DHS examined signatures of 
 the phase transition in $D_s(T)$ and $d D_s(T)/d T$ for lattice 
 sizes from $4 \times 4$ up to $64 \times 64$ and for temperatures
 as low as 87 K (taking the bandwidth to be 1 eV).  

  To study the effect of Cooper-pair fluctuations on the single-particle
  excitation spectrum, we examine the fully renormalized propagator and 
  self-energy obtained from numerical solutions of the equations for the FEA
  with $\phi_B = 0$.
  A detailed description of the FEA together with a summary of our 
  computational methods can be found elsewhere (Serene and Hess 1991,
  Deisz, {\em et al.} 1994, Deisz, {\em et al.} 1998a, Deisz, {\em et al.} 1998b).  
  Central to our discussion of the single-particle excitation spectrum 
  is Dyson's equation, which relates the fully renormalized propagator $G$ 
  to the self-energy $\Sigma$ and to the dispersion relation 
  $\epsilon_{\bf k}$ of the noninteracting system, 
   \begin{equation}
   G^{-1}({\bf k},  i\varepsilon_n) =
   \{ i\varepsilon_n - \xi_{\bf k} - \Sigma({\bf k}, \varepsilon_n) \},
   \end{equation}
   where $\xi_{\bf k} = \epsilon_{\bf k} - \mu$, $\mu$ is the chemical
   potential and $\varepsilon_n = (2n+1) \pi T$ are Matsubara frequencies.
   The FEA self-energy includes, in addition to the Hartree term,
   the second-order term and the exchange of spin-fluctuations, 
   density-fluctuations and Cooper-pair fluctuations, 
\begin{equation}
\Sigma({\bf r},\tau) = U^2 \left[ \chi_{ph}({\bf r},\tau) + T_{\rho\rho}
({\bf r},\tau) + T_{sf}({\bf r},\tau) \right] \,G({\bf r},\tau) +
U^2 T_{\rm pp}({\bf r},\tau)G(-{\bf r},-\tau).
\label{fea}
\end{equation}
    Here $T_{\rho\rho}$ and $T_{sf}$ are density and spin-fluctuation 
 $T$-matrices, and $T_{\rm pp}$ is the Cooper-pair fluctuation $T$-matrix. 
 For $U<0$ and sufficiently low temperature, $T_{\rm pp}$ provides the 
 largest contribution to the self-energy; 
 it is given by
\begin{eqnarray}
T_{\rm pp}({\bf q},\omega_m) & = &  \ \
  {{U \chi_{\rm pp}({\bf q},\omega_m)^2 }\over
  {1 + U \chi_{\rm pp}({\bf q},\omega_m)}},
\label{t_matrix}
\end{eqnarray}
where $\chi_{\rm pp}({\bf r}, \tau) = G({\bf r}, \tau) G({\bf r}, \tau)$.
We have included all contributions to the FEA self-energy
since this leads to a self-energy that is more accurate at high frequency
(Deisz, {\em et al.} 1997).

In Fig. 2, we show the superfluid density 
calculated for $U=-4$ and a density $n=0.75$ on a $64 \times 64$ and on a
$128 \times 128$ lattice.\footnote{We measure
all energies and temperatures in units of the hopping matrix element $t$.}
The sharp upturn in $D_s(T)$ signals the transition to superconductivity;
the finite size of the lattice is evident in the changes in $D_s(T)$
even for these large lattice sizes (Deisz {\em et al.} 1998a).  
As shown in the inset, the chemical 
potential becomes increasingly negative with increasing temperature 
as expected for a self-consistent conserving approximation that includes
particle-particle `ladder diagrams' (Serene 1989), 
as observed in quantum Monte Carlo calculations (Randeria {\em et al.} 1992), 
and {\em in contrast} to calculations in the `ladder 
approximation' that are {\em not self-consistent} 
(Schmitt-Rink {\em et al.} 1989).

 \begin{figure}[h]
  \centerline{\resizebox{4.0in}{4.0in}{\includegraphics{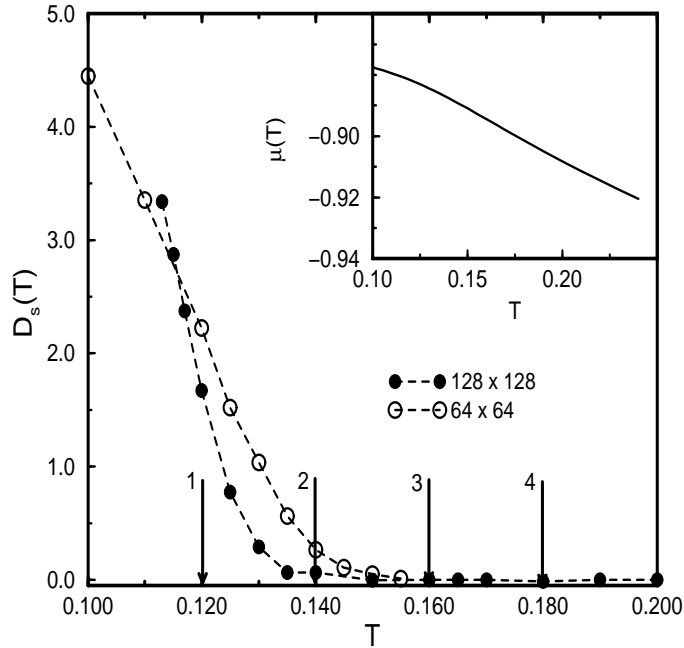}}}
\caption[]{{\footnotesize
  Superfluid density as a function of temperature for a
  $128 \times 128$ lattice ($\bullet$)  and $64 \times 64$ lattice
  ($\circ$). The arrows indicate the temperatures at which
  the excitation spectra are shown in panels 1 through 4 in Fig. 1.
  The inset shows the self-consistent chemical potential as
  a function of temperature. The (temperature independent) Hartree
  contribution to the self-energy is included in the chemical potential.
}}
 \end{figure} 

The single-particle spectral function $A({\bf k}, \varepsilon)$, provides a 
measure of the number of states accessible to an injected electron or hole.
Roughly speaking, the product of $A({\bf k}, \varepsilon)$ and a Fermi function
may be observed in angle-resolved photoemission experiments. 
The spectral function is related to the retarded propagator 
\begin{equation}
A({\bf k}, \varepsilon) = 
  - {{1}\over{\pi}} \; {\rm Im} \; G^{\rm R}({\bf k}, \varepsilon)
 = - {{1} \over {\pi}} \;
  {{{\rm Im} \; \Sigma^{R}({\bf k}, \varepsilon)} \over
  {(\varepsilon - \xi_{{\bf k}} - 
  {\rm Re} \; \Sigma^{R}({\bf k}, \varepsilon))^2
   + ({\rm Im} \; \Sigma^{R} ({\bf k}, \varepsilon))^2}}.
\end{equation}
We calculate $A({\bf k}, \varepsilon)$ from the retarded self-energy 
$\Sigma^{\rm R}({\bf k}, \varepsilon)$ which we obtain by analytic continuation 
from the imaginary frequency axis using 
Pad\'{e} approximants (Vidberg and Serene, 1977).  
In contrast to quantum Monte Carlo data, our self-energies have no 
numerical noise of statistical origin, but do contain  
roundoff errors dependent on the criterion for convergence; we converged our
self-energies so that the modulus of the largest change in $\Sigma$ 
for any $\bf k$ point or Matsubara frequency from one iteration to the 
next is less than $1.0 \times 10^{-7}$ for $64 \times 64$ lattices and less 
than $1.0 \times 10^{-8}$ for $128 \times 128$ lattices.  The density was
held fixed at $n=0.75$ to the same accuracy.

In Fig. 3a, we compare single-particle spectral functions $A({\bf k}, \varepsilon)$ 
for several temperatures at one point inside but very near the Fermi 
surface ${\bf k}_0 = (23,0) \pi /32$.  
The highest temperature shown ($T = 0.20$) for these calculations 
on a $128 \times 128$ lattice is well above any sign of 
superconductivity as shown in Fig. 2. At $T=0.20$, 
$A({\bf k}_0, \varepsilon)$ has a single rather broad peak.\footnote{ 
The large width of the peak and the temperature dependence of the 
width is not consistent with a quasiparticle of a Landau Fermi liquid. 
We do not pursue this point here.}
As the temperature is lowered in increments of $0.02$, the peak in 
$A({\bf k}_0, \varepsilon)$ shifts to lower energies; at the lowest
temperature shown ($T=0.12$), spectral weight is suppressed at 
$\varepsilon =0$ and it appears that the formation of a second peak
for $\varepsilon >0$ is incipient;  calculations on $64 \times 64$ 
lattices at lower temperatures clearly show 
a second peak for $\varepsilon >0$ (Hess {\em et al} 1996).  
Fig. 3b shows the real part of the denominator of the 
retarded Green's function, ${\rm Re} \; [G^{R} ({\bf k}_0, \varepsilon)]^{-1} =
\varepsilon - \xi_{{\bf k}_0} - {\rm Re} \Sigma^{R} ({\bf k}_0, \varepsilon)$.
The zero crossing of this expression, which corresponds to the energy of 
a quasiparticle excitation, occurs 
nearly at the peak in $A({\bf k}_0, \varepsilon)$. With decreasing
temperature the zero crossing shifts to lower energy. 
Structure in ${\rm Re} \; [G^{R} ({\bf k}_0, \varepsilon)]^{-1}$ 
at low energy also evolves with decreasing temperature.  For the 
$128 \times 128$ lattice, a perceptible superfluid density
is only evident for the lowest temperature (see Fig. 2). To better 
understand the evolution of the peak for $\varepsilon >0$ 
in $A({\bf k}_0, \varepsilon)$ with decreasing temperature, we 
also consider ${\rm Im} \; \Sigma^{R}({\bf k}_0, \varepsilon)$. 
As seen in Fig.  3c, structure evolves in
${\rm Im} \; \Sigma^{R}({\bf k}_0, \varepsilon)$ with decreasing 
temperature that is more complex than that which evolves in 
${\rm Re} \; [G^{R} ({\bf k}_0, \varepsilon)]^{-1}$. An examination of Fig. 3a 
shows that these changes do not lead to significant changes in the shape of
the spectral function for the higher temperatures where the only apparent
change in ${\rm Re} \; [G^{R} ({\bf k}_0, \varepsilon)]^{-1}$ is the shift of
the zero crossing to lower energy.  A more careful examination shows that
while changes in the energy dependence of ${\rm Im} \; \Sigma$ are reflected
in $A({\bf k}_0, \varepsilon)$, $|{\rm Im} \; \Sigma|$ is still rather large
and changes of ${\rm Re} \; [G_0^{R} ({\bf k}_0, \varepsilon)]^{-1} - 
{\rm Re} \Sigma ({\bf k}_0, \varepsilon)$ dominate.  Included in these are changes
in the chemical potential required to hold the density fixed.
Fig. 3d shows the spectral weight at $\varepsilon = 0$ as the temperature is 
decreased; a reduction of some $20 \%$ is evident as $T_c$ is approached 
and in the absence of any signature of superconductivity.  

 \begin{figure}[h]
 \centerline{\resizebox{5.0in}{4.0in}{\includegraphics{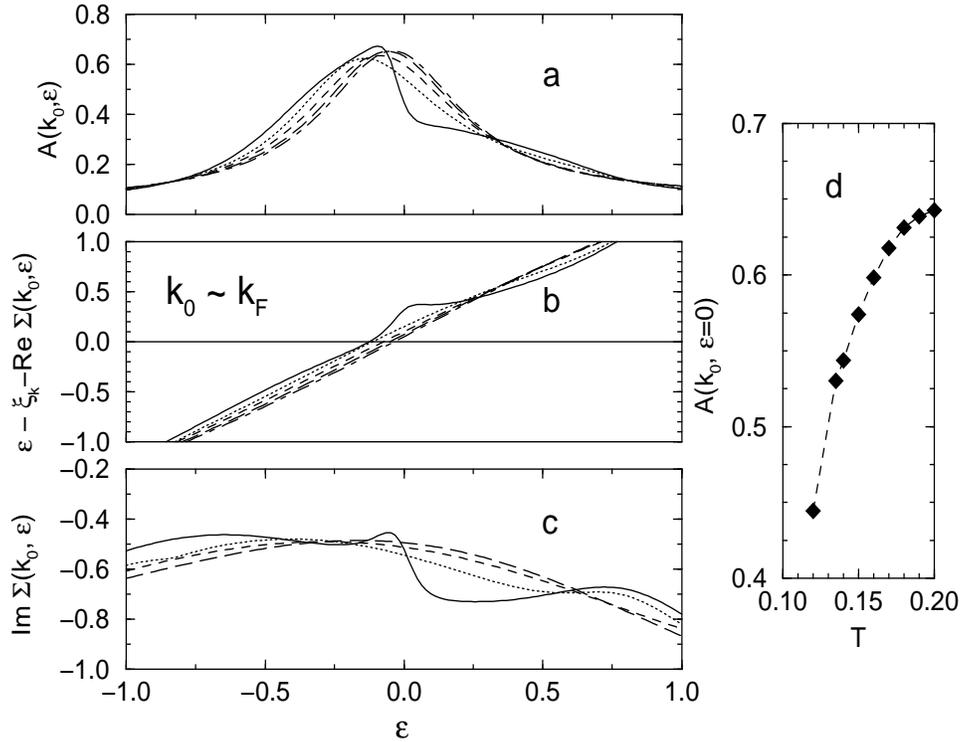}}}
\caption[]{{\footnotesize
 Contributions to the spectral function for
 ${\bf k}_0 = (23,0) \pi/32 $, a point near the Fermi surface
 shown for $T = 0.20$ (dash-dot), $0.18$ (long-dash), $0.16$ (dash),
 $0.14$(dotted), and $0.12$ (solid): (a) single-particle spectral weight,
 (b) Real part of the denominator of the fully renormalized retarded 
 Green's function
$\varepsilon - \xi_{\bf k} - {\rm Re} \Sigma^{R}({\bf k},
\varepsilon)$, and (c) Imaginary part of the self-energy.
Also shown in (d) is the spectral weight at zero energy as a
function of temperature.  Comparison with the superfluid density
and it's derivative in Fig. 2 suggests that the slight kink around
$T=0.135$ is another signature of the phase transition to superconductivity.
}}
\end{figure}

The total DOS is a sum of $A({\bf k}, \varepsilon)$ over all momenta in
the zone; our results suggest that the DOS will show a similar suppression 
of spectral weight. The suppression will differ quantitatively from
that shown in Fig. 3d, which tracks the peak of what 
might be regarded as the `coherent part' of the (non-Fermi liquid) 
quasiparticle excitation. Changes from incoherent parts of spectral 
functions for other ${\bf k}$ are included in the DOS. Depending in 
detail on how Im $\Sigma({\bf k}, \varepsilon=0)$ changes with
temperature for ${\bf k}$ near ${\bf k}_{F}$, these may lead to additional 
reduced contributions to the DOS as coherent quasiparticle
excitations or possible gap structure in the low-energy excitation 
spectrum attempt to evolve. The evolution of these `dimples' 
in the spectral function for ${\bf k}$ near ${\bf k}_{F}$
as $\varepsilon \rightarrow 0$ would be analogous to those observed 
as a coherent state evolves in the Anderson lattice model 
(McQueen {\em et al.} 1993). 

These changes in single-particle excitation spectra are accompanied by 
the self-consistent formation of a sharp peak in 
$T_{\rm pp} ({\bf q}, \omega_m)$ at ${\bf q} = 0$ and $\omega_m =0$ which
evolves as an instability is approached (see Fig. 3 of DHS).  For larger
lattices, $T_{\rm pp}$ is more sharply peaked with a larger
$T_{\rm pp} ({\bf 0},0)$.
Supposing that this peak leads to the dominant contribution to the 
self-energy,\footnote{A similar model was considered by Patton
(1971).} we find, 
\begin{eqnarray}
\Sigma({\bf k}, i\varepsilon_n) & = & {{T} \over {N}} \sum_{m, {\bf q}}
  G(- {\bf k} + {\bf q}, -\varepsilon_n + \omega_m)  
  T_{\rm pp} ({\bf q},\omega_m) \\ \nonumber
  & \approx & \tilde{t}_{pp} G(- {\bf k}, -\varepsilon_n), 
\end{eqnarray}
 where $\tilde{t}_{pp} = T/ N \sum_{|{\bf q}|< \xi} T_{\rm pp} ({\bf q},0)$,
 $\xi$ is an appropriate correlation length,  and $N$ is the number 
 of sites.  Inserting this result into Dyson's equation 
 leads to a connection between $G$ at $({\bf k},\varepsilon_n)$ and 
 the time reversal conjugate point 
 $(- {\bf k}, -\varepsilon_n)$ as expected for a pairing interaction.
 Taking ${\bf k}$ to be on the Fermi surface and 
 using Dyson equations for $G$ for two $({\bf k},\varepsilon_n)$ points
 related by time reversal conjugation, we find for the single-particle 
 spectral function,
 \begin{equation}
   A({\bf k}_F,\varepsilon) = {{1}\over{2 \pi \tilde{t}_{\rm pp}}} 
   \sqrt{4\tilde{t}_{\rm pp} - \varepsilon^2}; \; \; 
   |\varepsilon| < 2 \sqrt{\tilde{t}_{\rm pp}}. 
 \end{equation}
  This is a parabolic shaped spectral function centered on $\varepsilon =0$
  with a maximum value $\propto 1/\sqrt{\tilde{t}_{\rm pp}}$ and with all spectral 
  weight contained in the interval $|\varepsilon| < 2 \sqrt{\tilde{t}_{\rm pp}}$.
  As $\tilde{t}_{\rm pp}$ increases, the maximum is reduced and the 
  size of the interval increases so that the sum of all spectral weight at 
  any given ${\bf k}$-point is unity.  This is not the entire contribution to the
  self-energy, but this contribution is rapidly changing, leading to a 
  suppression of spectral weight for any ${\bf k}$-point at the Fermi surface.
  We note in passing that  spectral weight is pushed away from the Fermi 
  surface, and for ${\bf k}$-points off of the Fermi surface, a second 
  peak appears in the spectral function consistent with particle-hole 
  coherent Bogoliubov-like excitations.
  While this model is too simplistic to correctly capture the essential
  features of the full calculation, it does illustrate how a superconducting
  transition in two dimensions that separates two disordered phases 
  {\em might} occur in the FEA without giving rise to a finite gap but 
  still leading to a suppression of low-energy spectral weight in the 
  normal state.

    We return now to the excitation spectra shown in Fig. 1 for ${\bf k}$ along the 
  $(1,0)$ direction in the zone, which  
  shows that the quasiparticle excitations are still evolving as
  the superconducting transition is approached,  intermingled in a 
  self-consistent way with the evolution of superconducting fluctuations. 
  As the temperature is lowered, the suppression of spectral weight at zero
  energy is evident in the failure of a sharp peak to emerge in the narrow 
  green region sandwiched between much sharper (red) quasiparticle and 
  quasihole peaks slightly above and below zero energy.  
  The largest value of the spectral functions occurs at the 
  lowest temperature shown and is $\sim 1.4$.  We have noted that
  the spectral weight at the point closest to the Fermi surface decreases strongly
  below $T \sim 0.18$ and falls by some $20\%$ just above $T_c$.  So, relative 
  to the sharp quasiparticle excitations above and below the Fermi surface, 
  spectral weight at the peak of the quasiparticle excitation near the Fermi surface
  is suppressed by roughly $50\%$.  At the lowest temperature, 
  a finite superfluid density exists (see Fig. 2), finite but small spectral weight
  exists at zero energy, and hints of additional structure emerging from the
  region around zero-energy are evident. These ``satellite'' peaks suggest the 
  coherent coupling of particle and hole excitations expected for a superconducting
  state, Bogoliubov excitations. 
  
  While we have explicitly shown that spectral weight at low-energy is suppressed
  for this sizeable density,
  the quantitative result is less than observed in experiments for BSCCO which
  show a large reduction in spectral weight.  The quantitative dependence 
  of the spectral weight just above $T_c$ on the interaction strength and
  density is so far unknown.  The dependence on the symmetry of the order
  parameter is also a subject of current research (Engelbrecht 1997).
  It is also largely unknown how Cooper-pair fluctuations 
  interact with other possible mechanisms for the formation of the pseudogap
  that are not a consequence of superconductivity. 
  Further work seeking to elucidate these issues is underway.

    We have presented fully self-consistent calculations of the temperature 
  dependent superfluid density together with the single-particle excitation spectrum
  for a strong coupling superconductor with an attractive interaction of half 
  the bare bandwidth in two dimensions on large $128 \times 128$ lattices.  
  We have taken the electronic density to be roughly that expected from electronic
  structure calculations. Our  
  calculations show the evolution of quasiparticle excitations self-consistently 
  intermingled with superconducting fluctuations, and explicitly demonstrate a 
  significant suppression of spectral weight at low energy as the superconducting 
  transition is approached from {\em above}.
  For temperatures just below the transition temperature, a gap is not evident.
  A simplistic self-consistent model was presented to illustrate how the FEA might
  produce a phase transition between disordered states in two dimensions without
  producing a finite gap in the single particle excitation spectrum.

  \begin{center}
  {\large  Acknowledgements}
 \end{center}
 
  DWH acknowledges the support of the Office of Naval Research, 
  and J.J.D. acknowledges the support in part of NSF Grant ASC-9504067.
  This work was also supported by a grant of computer time from the 
  DoD HPC Shared Resource Centers: 
  Naval Research Laboratory Connection Machine facility CM-5; 
  Army High Performance Computing Research Center under the 
  auspices of Army Research Office contract number 
  DAAL03-89-C-0038 with the University of Minnesota.
  DWH would like to thank J. Erwetowski for a history of
  supurb administrative support.

 \begin{center}
  {\large References}
 \end{center}

\begin{list}{ }{\leftmargin 0.0in \itemsep 0.0in}

   \item Baym, G., 1962, {\it Phys. Rev. B}, {\bf 127}, 1391.

   \item Bickers, N.E., Scalapino, D.J., and White, S.R., 1989,
     {\it Phys. Rev. Lett.} {\bf 62}, 961.

   \item Deisz, J.J., Hess, D.W., and Serene, J.W., 1994,
    {\it Recent Progress in Many-Body Theories}, vol. 4,
    edited by E. Schachinger, {\it et al.}, Plenum, New York.

   \item Deisz, J.J., Hess, D.W., and Serene, J.W., 1997, 
     {\it Phys. Rev. B} {\bf 55}, 2089. 

   \item Deisz, J.J., Hess, D.W., and Serene, J.W., 1998a, 
    {\it Phys. Rev. Lett.} {\bf 80}, 373.

   \item Deisz, J.J., Hess, D.W., and Serene, J.W., 1998b, 
      in preparation.

   \item Ding, H., Norman, M.R., Yokoya, T., Takeuchi, T., Randeria, M.,
    Campuzano, J.C., Takahashi, T., Mochiku, T. and Kadowaki, K., 1997,
    {\it Phys. Rev. Lett.} {\bf 78}, 2628.

   \item Doniach, S., and Inui, M., 1990, {\it Phys. Rev. B} {\bf 41}, 6668.

   \item Eagles, D.M., 1969, {\it Phys. Rev.} {\bf 186}, 456. 

   \item Emery, V.J., and Kivelson, S.A., 1995, {\it Nature} {\bf 374}, 434.

   \item Engelbrecht, J., Nazarenko, A., Randeria, and M., Dagotto, E., 1997, 
    cond-mat/9705166.

   \item Fisher, M.E., Barber, M.N., and Jasnow, D., 1973, 
     {\it Phys. Rev. A} {\bf 8}, 1111.

   \item Hess, D.W., Deisz, J.J., and Serene, J.W., 1996, unpublished.

   \item Leggett, A.J., 1980, in {\it Modern Trends in Condensed Matter},
    A. Pekalski and R. Przystawa, eds. (Springer-Verlag, Berlin); and
    {\it J. Phys. (Paris) Colloq.}, {\bf 41}, C7-19.

   \item Luo, J., and Bickers, N.E., 1993, {\it Phys. Rev. B}
     {\bf 48}, 15983.

   \item McQueen, P.G., Hess, D.W., and Serene, J.W., 1993, {\it Phys.
    Rev. Lett.} {\bf 71}, 129.

   \item Micnas, R., Ranninger, J., and Robaszkiewicz, S., 1990,
     {\it Rev. Mod. Phys.} {\bf 62}, 113. 

   \item Nozi\`{e}res, P., and Schmitt-Rink, S., 1985, 
     {\it J. Low Temp. Phys.} {\bf 59}, 195. 

   \item Patton, B.R., 1971, Phys. Rev. Lett. {\bf 27}, 1273.

   \item Randeria, M., Duan, J., and Shieh, L., 1989, 
     {\it Phys. Rev. Lett.} {\bf 62}, 981.

   \item Randeria, M., Trivedi, N., Moreo, A., and Scalettar, R., 1992,
   {\it Phys. Rev. Lett.} {\bf 69}, 2001.

   \item S\'{a} de Melo, C.A.R., Randeria, M., and Engelbrecht, J., 1993,
   {\it Phys. Rev. Lett.} {\bf 71}, 3202.

   \item Scalapino, D.J., White, S.R., and Zhang, S.C., 1992,
    {\it Phys. Rev. Lett.} {\bf 68}, 2830.

   \item Serene, J.W., 1989, {\it Phys. Rev. B} {\bf 40}, 10873.

   \item Serene, J.W., and Hess, D.W., 1991, {\it Phys. Rev. B} {\bf 44}, 3391.

   \item Schmitt-Rink, S., Varma, C.M., and Ruckenstein, A.E., 1989, 
         {\it Phys. Rev. Lett.} {\bf 63}, 445.

   \item Trivedi, N., and Randeria, M., 1995, {\it Phys. Rev. Lett.} 
	 {\bf 75}, 312.

  \item Vidberg, H.J., and Serene, J.W., 1977, {\em J. Low Temp. Phys.},
         {\bf 19}, 179.

  \item Yang, C.N., 1962, {\it Rev. Mod. Phys.} {\bf 34}, 694.

\end{list}

\end{document}